\documentclass{cpcauth}
\usepackage[dvips]{graphics,psfrag}
\usepackage{epsfig}
\usepackage{psfig}
\usepackage{colordvi}

\newcommand{\bq}{\begin{equation}}
\newcommand{\eq}{\end{equation}}
\newcommand{\bqa}{\begin{eqnarray}}
\newcommand{\eqa}{\end{eqnarray}}

\newcommand{\bit}{\begin{itemize}}
\newcommand{\eit}{\end{itemize}}
\newcommand{\nn}{\nonumber}

\newcommand{\Tr}{ {}{\rm Tr}{} }

\newcommand{\ReTr}{ {}{\rm ReTr}{} }

\newcommand{\unit}{1\hspace*{-1.5mm}\mbox{l}}

\begin{document}
\begin{frontmatter}
\title{
\vspace*{-80pt}
{\normalsize
\mbox{} \hfill BI-TP 2001/31\\
\mbox{} \hfill November 2001}\\
\vspace*{50pt}
Common features of deconfining and chiral critical points in QCD and the
  three state Potts model in an external field }
\author{F. Karsch, Ch. Schmidt and S. Stickan}
\address{Fakult\"at f\"ur Physik, Universit\"at Bielefeld, D-33615 Bielefeld}

\begin{abstract}
In the presented study we investigated the second order endpoints of the lines of first order
phase transitions which emerge for the QCD in the heavy and light quark mass
regime and for the three-dimensional three state Potts model with an external
field. We located the endpoints with Binder cumulants and constructed the
energy-like and ordering field like observables. The joint probability
distributions of these scaling fields and the values of the
Binder cumulant confirm that all three endpoints belong to the universality
class of the 3-dimensional Ising model.
\end{abstract}




\end{frontmatter}

\section{Introduction}
The spontaneous breaking of global symmetries at low and their restoration at high
temperature are common features of quantum field theories and
statistical models. It is well known that universal properties at finite
temperature phase transitions in (3+1) dimensional gauge theories are related
to those in 3-dimensional spin models \cite{svet}.

In the presented study we investigated
the proposed connection between the SU(3) gauge theory with two heavy
quark flavors and the three state Potts model with an external field \cite{KS} and in addition the
physically more relevant case with three light quark flavors \cite{KLS}. In all
three cases a first order phase transition line emerges and ends in a
second order endpoint \cite{GGP}.

To determine the universality class of these endpoints its precise
location is mandatory. Moreover, observables which are sensitive to the
universality classes are needed. They are constructed using a method originally proposed to
study the liquid gas transition point \cite{RM}. We extended this approach to
the more complex case of a quantum field theory.

\section{The Potts Model}
The 3-dimensional three state Potts model with an external field is
defined with the spin variables $\sigma_i\in\{1,2,3\}$ by
\bqa
{\mathcal H} &=&-\beta \sum_{\langle i,j
  \rangle} \delta(\sigma_i,\sigma_j) - h \sum_i \delta(\sigma_i,\sigma_g) \\
&\equiv&-\beta E -h M
\eqa
A non vanishing $h>0$ favors the magnetization in the direction of the spin
$\sigma_g$. The partition function on a finite lattice of size $L^3$ is then given by
\bqa
Z(\beta,h,L)&=&\sum_{\{\sigma_i\}} \exp(-\mathcal{H}).
\eqa
The pseudo-critical line $\beta_{pc,L}(h)$ is defined by the maxima of the susceptibility
\bqa
\chi_L &=& \frac{1}{L^3}\left( { \langle M^2\rangle - \langle M \rangle^2 }\right)\nn
\eqa
A finite size analysis of $\chi_L$ leads to  {$h_c\in [0.0005,0.0010]$}
\cite{KS} which is not sufficient for a precise location of the endpoint.

\section{Quantum Chromo Dynamics}
To simulate the QCD we used the standard staggered fermion action
\bqa
S_{F}(\psi,\bar{\psi},U,m_q)&=&\sum_{n,m}\bar{\psi}(n)K_{nm}^{KS}(U,m_q)\psi(m)
\eqa
with the fermion fields $\psi,\bar{\psi}$ and the link variables
${U_\mu(n)=\e^{-igaA_\mu(n)}}\!\!\in\!\! SU(3)$ in the staggered fermion matrix
\bqa
K_{mn}^{KS}(U,m_q)=\delta_{nm}m_q +\sum_\mu (-1)^{n_1+...+n_{\mu-1}} \nn\\
\times\frac{1}{2}\left({\delta_{n+\hat\mu,m}U_\mu(n)-\delta_{n,m+\hat\mu}U^\dagger_\mu(m)
  }\right).
\eqa
For the gauge action we choose the standard Wilson plaquette action
\bqa
S_G(U) &=& \sum_x
\sum_{1\leq\mu<\nu\leq4}\left\{{\unit-\frac{1}{3}\ReTr U_{\mu\nu} } \right\}.
\eqa
Integrating out the fermion fields we obtain the partition function on a $N_\sigma^3\times N_\tau$ lattice
\bqa
Z&=&\int\prod_{n,\mu} dU_\mu(n)\;\; \e^{-S^{QCD}}
\eqa
with
\bqa
S^{QCD}&=&\left\{ {-\beta S_G(U)+\frac{n_f}{4}\Tr\ln K^{KS}(U,m_q) } \right\} .
\eqa
and $n_f$ as the number of quark flavors. Expanding $\Tr\ln K = -\sum_l \frac
{\kappa^{l}} {l}\; \Tr M^l$ with $K=1-\kappa M$ and $\kappa=1/2m_q$ one can show that for heavy quarks and $N_\tau$=4
\bqa
S_F&=&\frac{n_f}{m^4_q}\sum_n\left( {L(n)+L^\dagger(n) }\right) +O(\kappa^6),
\eqa
i.e. the quark mass enters the action with $m_q^{-4}$. \\
The order parameters of the QCD phase transitions are the Polyakov loop
\bqa
\langle L \rangle &=& N_\sigma^{-3} \left\langle { \sum_{\bf n}
  \Tr\prod_{n_4=1}^{N_\tau} U_{\hat{\tau}}({\bf n},n_4) }\right\rangle
\eqa
in the limit $m_q\rightarrow\infty$ for the deconfinement transition and the
chiral condensate
\bqa
  \langle\bar\psi\psi\rangle&=&\frac{n_f}{4N_\sigma^3N_\tau}
  \left\langle{\Tr K^{-1}}\right\rangle
\eqa
in the limit $m_q\rightarrow 0$ for the chiral transition.

\section{The Simulation Parameters}
For the Potts model we used the Wolff cluster algorithm with a ghost spin in
order to implement the external field and to simulate on lattices of size
$V\!=\!40^3-70^3$. We define a new configuration after 1000
configurations to get autocorrelations of the energy $E$ between 7 and 25 close
to the pseudo critical points. We simulate at 3 to 4 different $\beta$-values
for every $h$ and perform 10000 updates at each ($\beta,h$)-pair of couplings. In
the analysis of the data we make use of the Multihistogram-Method \cite{FS} to
interpolate between $\beta$ and $h$ values.\\

The QCD simulations are performed with the Hybrid R algorithm. We used lattices
of size $8^3\times 4$, $12^3\times 4$, $16^3\times 4$ and for the heavy quark
regime an additional size of $24^3\times4$. We simulated with three mass
degenerate quark flavors of masses $m_qa\!=\!0.03, 0.0325, 0.035, 0.04$ close to the chiral
and two mass degenerate quark flavors of $m_qa\!=\!1.5,1.7,1.8,2.0$ close to the
deconfinement transition. For 3-4 different $\beta$-values per mass
parameter we perform $(1-3)/(3-7)\times 10^4$ updates
at small/large bare quark masses.  In the analysis a reweighting in the
$\beta$-direction has been used.

\section{Constructing the Scaling Fields}

In order to construct observables for the Potts model which are more sensitive
to the universality class of the endpoint we make the assumption
\bqa
{\mathcal M}=M+sE,\;\;\;&&\;\;\; {\mathcal E}=E+rM
\eqa
and rewrite the Hamiltonian in terms of these new fields
\bqa
{\mathcal H}&=&-\tau{\mathcal E}-\xi{\mathcal M}
\eqa
where the new couplings are given by
\bqa
\xi=\frac{1}{1-rs}(h-r\beta),\;\;\;&&\;\;\;\tau=\frac{1}{1-rs}(\beta-sh).
\eqa
Identifying $\tau$ with the direction of
the pseudocritical line at the endpoint and demanding that the
energy-like fluctuations and those of the ordering field are uncorrelated we
fix the parameter $r$ and $s$ by
\bqa
r^{-1}=\left. {\frac {d\beta_c(h)}{dh} } \right|_{h=h_c},\;\;\; &&\;\;\;
\langle\delta{\mathcal M}\delta{\mathcal E}\rangle=0
\eqa
where $\delta{\mathcal X}\equiv {\mathcal X}-\langle {\mathcal X} \rangle$.
In QCD we define analogously for the light quarks
\bqa
{\mathcal M}=\bar\psi\psi+sS_G,\;\;\; && \;\;\;{\mathcal E}=S_G+r\bar\psi\psi
\eqa
and similarly
\bqa
{\mathcal M}=L+sS_G,\;\;\; &&\;\;\; {\mathcal E}=S_G+rL
\eqa
for heavy quark masses. Rewriting the action $S^{QCD}$ in terms of these
variables is not possible, but the new energy-like and ordering field-like
variables ${\mathcal E}$ and ${\mathcal M}$ define an effective Hamiltonian
\bqa
{\mathcal H}_{eff}=\tau{\mathcal E}+\xi{\mathcal M}
\eqa
which will control the universal behavior in the vicinity of the critical
points in QCD.
We similarly make a linear ansatz for the new couplings
\bqa
\tau&=&(\beta-\beta_c)+A(h-h_c),\\
\xi&=&(h-h_c)+B(\beta-\beta_c)
\eqa
with $h\!=m\;(m^{-4})$ for light (heavy) quark
masses. We use the constraints
\bqa
0=\left. {\frac{\partial\langle{\mathcal M}\rangle_{{\mathcal H}_{eff}}}{\partial \tau}
  }\right|_{\tau=\xi=0}=\langle\delta{\mathcal M}\;\delta{\mathcal E} \rangle_{S^{QCD}}
\eqa
and
\bqa
B=-\left. {\frac{\partial h_c(\beta)}{\partial\beta} } \right|_{\beta=\beta_c}
\eqa
to fix three of the four constants $r$,$s$ and $A$,$B$.

Using a scaling ansatz $r_L\!=\!r_\infty\!+\frac{c}{L^3}$ we found for the Potts
model $r_{\infty}\!=\!-0.690(2) \approx -s$. Thus the transformation is just a rotation.
In the case of the QCD a size dependence of $r$ and $s$ is observed but a quantitative
scaling analysis was not feasible. We therefore used results from the largest
lattice. Close to the endpoint we define for the chiral region
  $(s,r)=(0.41\pm 0.52,0.550\pm 0.007)$ and for the deconfinemt transition $(s,r)=(0.67\pm 2.48,-0.223\pm0.043)$.


\section{Locating the Endpoints}

\begin{figure}[b]
  \hskip -2.8mm
  \epsfig{file=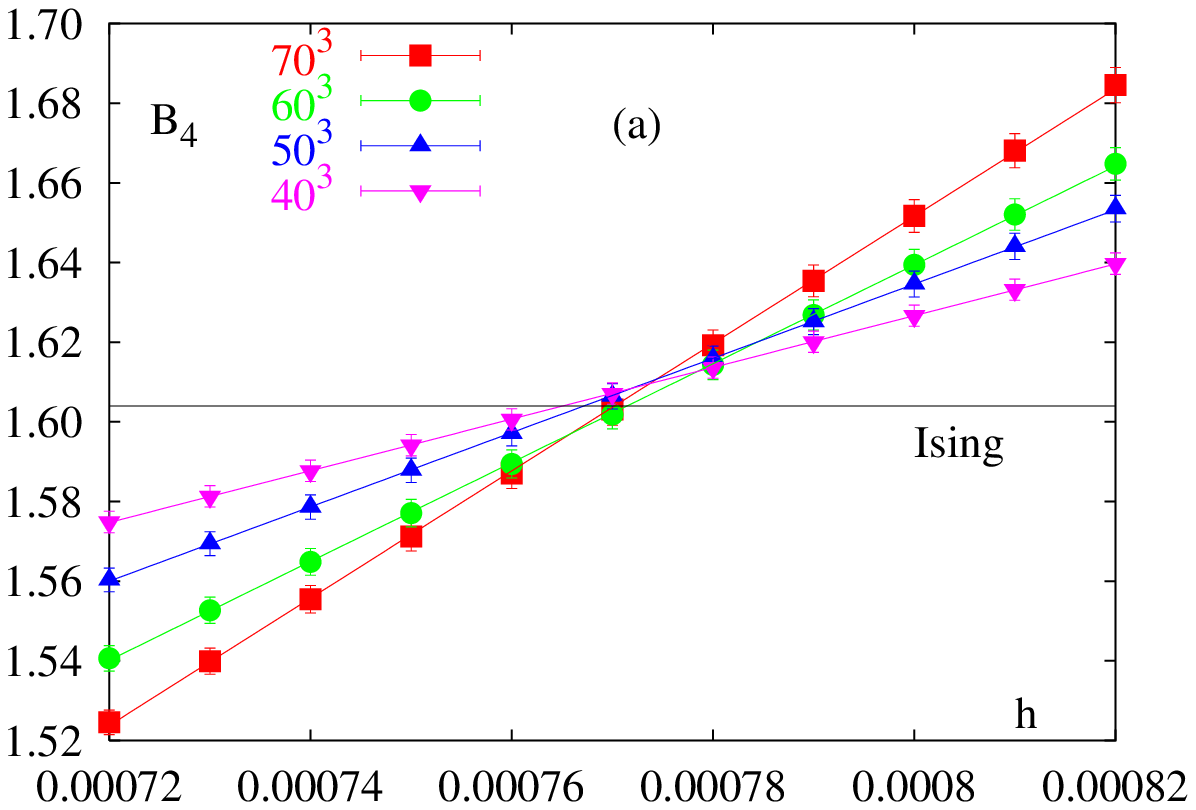,width=200\unitlength,angle=0}
  \epsfig{file=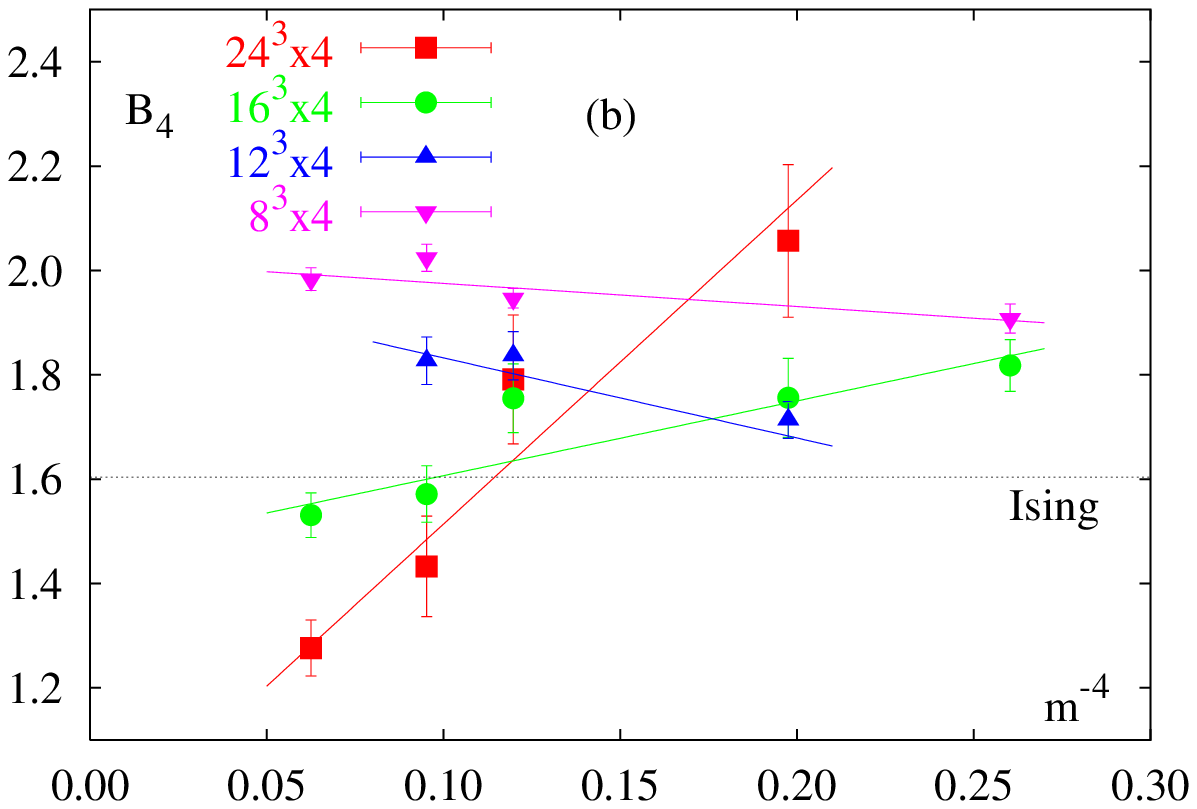,width=200\unitlength,angle=0}
  \epsfig{file=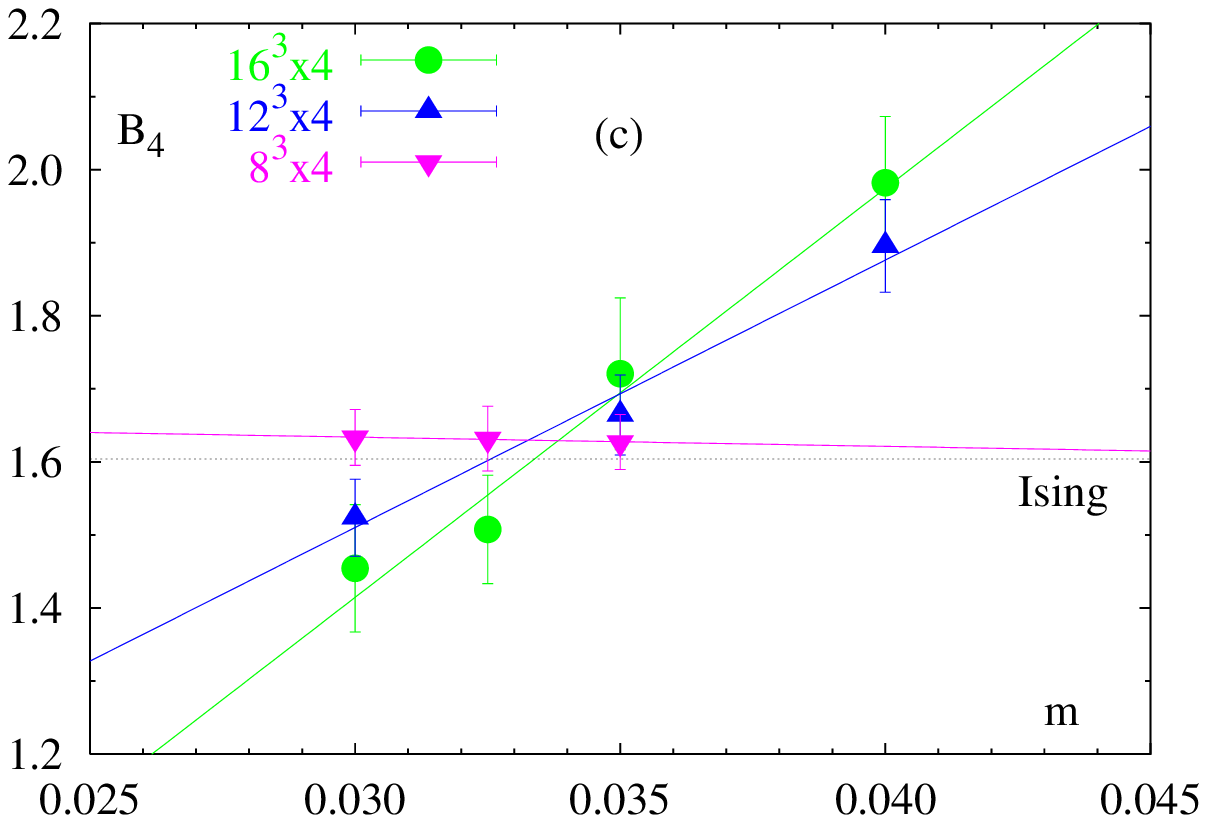,width=200\unitlength,angle=0}
  \caption{ $B_{4,L}$ of the Potts model (a) and for the QCD with heavy (b)
  resp. light (c) quark masses.(Lines are drawn to guide the eyes)}
  \label{cumulant-picture}
\end{figure}

For a precise location of the critical points we use the Binder cumulants
\bqa
B_{3,L} = \frac{ \left\langle  { (\delta{\mathcal M})^3 } \right\rangle_L } {
  \left\langle  { (\delta{\mathcal M})^2 } \right\rangle^{3/2}_L },\;\;\; && \;\;\;
B_{4,L} = \frac{ \left\langle  { (\delta{\mathcal M})^4 } \right\rangle_L }
{ \left\langle  { (\delta{\mathcal M})^2 } \right\rangle^{2}_L }.
\eqa
The cumulant $B_{3,L}$ should vanish for $\beta\!\!=\!\!\beta_{pc}$ and the cumulant
$B_{4,L}$ should take on a volume independent value $B_{4,c}$ at the
endpoint. This value is 1.602(2), 1.242(2) or 1.092(3) for the 3-dimensional
Ising, O(2) or O(4) universality classes,  respectively. We found for the Potts model
\cite{KS} a value of 1.609(4)(10) at (0.54938(2),0.000775(10)) where the first error is a
statistical error and the second one is due to uncertainties in locating
the endpoint. For the QCD we found \cite{KLS} a value of 1.547(46) at
(5.6756(5),1.74(4)) and 1.639(24) at (5.1458(5),0.033(1)) in the heavy and
light quark mass regimes respectively (Fig.\ref{cumulant-picture}). This
strongly suggest that all three endpoints fall in the Ising universality class. \\

\section{Joint Probability Distributions}

\begin{figure}[b]
  \vskip -15mm
  \psfrag{E}[c]{$\scriptstyle\Lambda_{\mathcal E}^+\delta{\mathcal E}$}
  \psfrag{M}[c]{$\scriptstyle\Lambda_{\mathcal M}^+\delta{\mathcal M}$}
  \epsfig{file=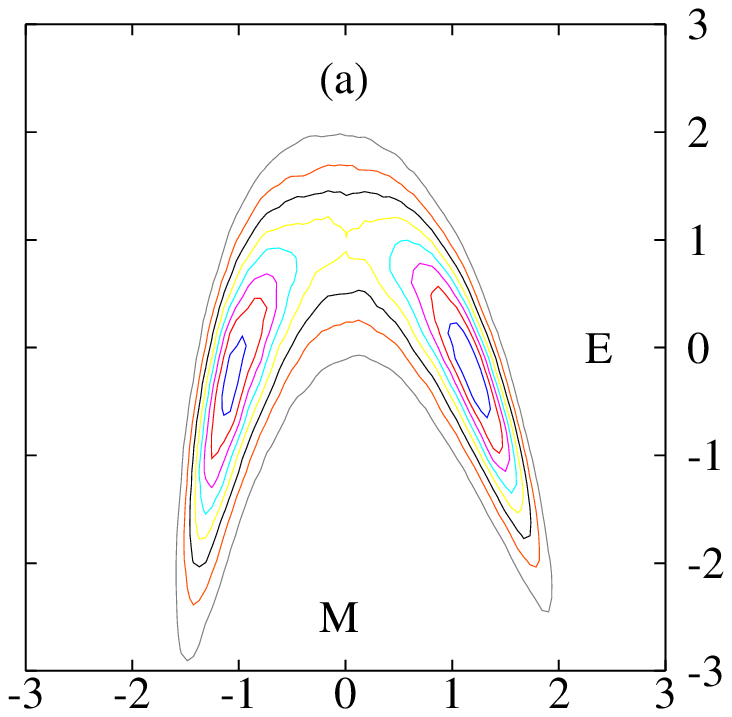,width=200pt,angle=0}
  \vskip -30mm
  \epsfig{file=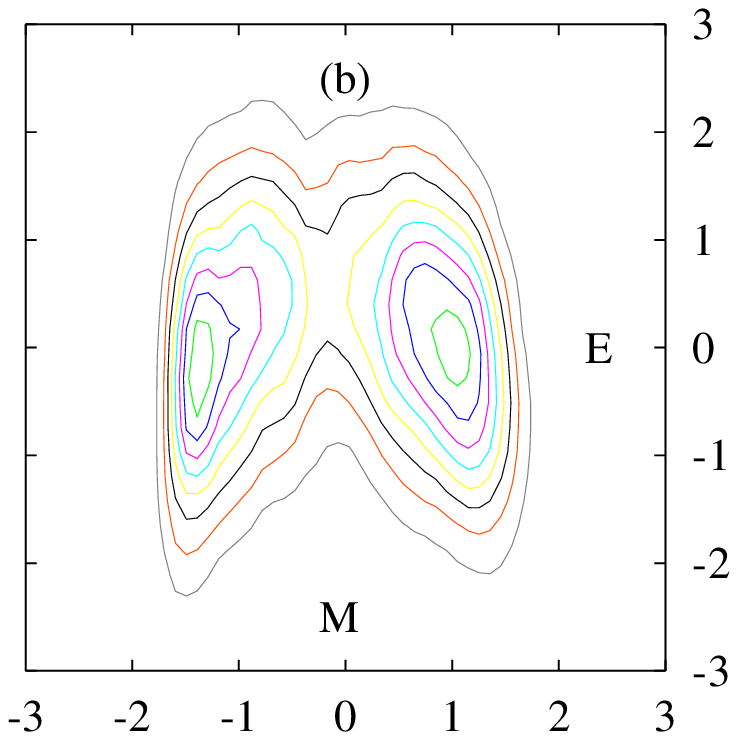,width=200pt,angle=0}
  \vskip -30mm
  \epsfig{file=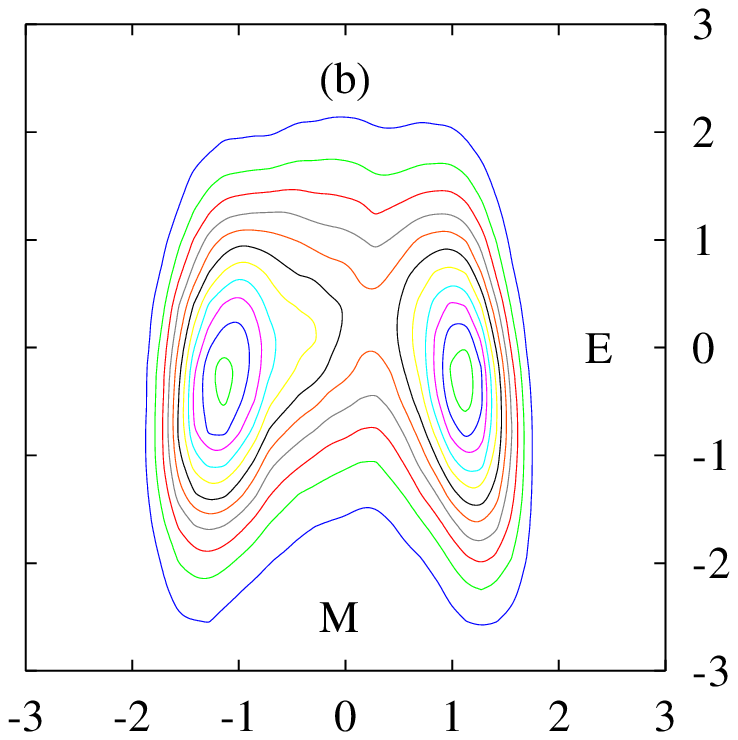,width=200pt,angle=0}
  \vskip -15mm
  \caption{ Histogram $\hat p_{{\mathcal E}{\mathcal M}}$ of the Potts model (a)
    at the endpoint and of the QCD with $m=1.8$ (b), $m=0.325$ (c) close to the
    end points.}
  \label{Histograms}
\end{figure}

With the finite size scaling ansatz \cite{WB} for joint probability distributions of ${\mathcal E}$ and ${\mathcal M}$
\bqa
p_L({\mathcal E},{\mathcal M})\propto \tilde p_{\mathcal EM}\left( {
  \Lambda_{\mathcal E}^+\delta{\mathcal E},\Lambda_{\mathcal
    M}^+\delta{\mathcal M},\Lambda_{\mathcal E} \tau,\Lambda_{\mathcal M}\xi } \right)
\eqa
with  $\Lambda_{\mathcal E}\!\!=\!\!a_{\mathcal E}L^{1/\nu}$,$\Lambda_{\mathcal M}\!\!=\!\!a_{\mathcal M}L^{d-\beta/\nu}$ and $\Lambda_{\mathcal M}\Lambda_{\mathcal M}^+\!\!=\!\!\Lambda_{\mathcal E}\Lambda_{\mathcal E}^+\!\!=\!\!V $  one obtains the universal function
\bqa
p_L({\mathcal E},{\mathcal M})\equiv\Lambda_{\mathcal E}^+\Lambda_{\mathcal M}^+ \hat p_{\mathcal
  EM} \left( { \Lambda_{\mathcal E}^+\delta{\mathcal E},\Lambda_{\mathcal M}^+\delta{\mathcal M} } \right)
\eqa
at ${\xi\!=\!\tau\!=\!0}$ (Fig.\ref{Histograms}). Comparing these distributions
with those obtained for other universality classes
we again conclude that the endpoint belongs to the Ising universality class.

\end{document}